\documentclass[twocolumn]{pasj00}
\Received{$\langle$reception date$\rangle$}
\Accepted{$\langle$acception date$\rangle$}
\Published{$\langle$publication date$\rangle$}
\SetRunningHead{Astronomical Society of Japan}{Suzaku observation of G\,359.77$-$0.09}

\usepackage{times}

\begin{document}

\title{A Super Bubble Candidate in the Galactic Center and
a Local Enhancement G\,359.77$-$0.09}
\author{Hideyuki \textsc{Mori},\altaffilmark{1,2} 
Yoshiaki \textsc{Hyodo},\altaffilmark{1} 
Takeshi Go \textsc{Tsuru},\altaffilmark{1} 
Masayoshi \textsc{Nobukawa},\altaffilmark{1} 
and Katsuji \textsc{Koyama},\altaffilmark{1}}
\altaffiltext{1}{Department of Physics, Graduate School of Science,
Kyoto University, Sakyo-ku, Kyoto, 606-8502}
\altaffiltext{2}{Department of High Energy Astrophysics, Institute of
Space and Astronautical Science (ISAS), Japan Aerospace Exploration
Agency (JAXA), 3-1-1, Yoshinodai, Sagamihara, Kanagawa, 229-8510}
\email{mori@astro.isas.jaxa.jp}

\maketitle

\begin{abstract}
A $20' \times 16'$ elliptical ring-like structure has been found near
the Galactic center in the narrow energy band corresponding to the
K$\alpha$ line from He-like sulfur.  In the ring, two diffuse sources
are found, a supernova remnant candidate G\,359.79$-$0.26 and an
unidentified source G\,359.77$-$0.09. The X-ray spectrum of
G\,359.77$-$0.09 is similar to that of G\,359.79$-$0.26, which can be
explained by an absorbed thin thermal plasma model with temperatures
of $0.7$ and $1.0$~keV.  The absorption column densities of these two
sources are large ($N_{\rm H} = 6.9 \times 10^{22}$ and $4.5 \times
10^{22}$~cm$^{-2}$) and are consistent with that of the Galactic
center distance.  The X-ray spectrum extracted from the ring-like
structure is also represented by an absorbed thin thermal plasma model
($kT_{\rm e} \sim 0.9$~keV).  The thermal energy of the plasma filling
the ring-like structure is estimated to be $1.0 \times 10^{51}$~erg.
We therefore propose that the two sources comprise a single ring-like
object, which is possibly a super bubble with a size of $49$~pc
$\times$ $40$~pc in the Galactic center region.
\end{abstract}

\section{Introduction}
\label{section:introductin}
Super bubbles (SBs) are large-scale diffuse objects in galaxies heated
by shocks of sequential supernova (SN) explosions and/or strong
stellar winds from a number of massive stars. The temperature of the
shock-heated gas reaches $10^{6-7}$~K so that SBs emit diffuse soft
X-rays.  The most complete and measurable samples of the X-ray SBs are
those in the Large Magellanic Cloud (LMC).
\citet{2001ApJS..136..119D} surveyed the X-ray SBs in the LMC with the
ROSAT satellite and reported X-ray emission from 13 SBs.  However, no
detailed spectral information was available due to limited photon
statistics.  The XMM-Newton satellite re-visited some of these SBs and
found their temperatures to be typically 0.1--0.2~keV, similar to
those of old supernova remnants (SNRs, e.g.,
\cite{2004A&A...418..841N}).  One exception is 30 Dor C where some
parts of the SB show hard spectra with a power-law index of 2.7
\citep{2004ApJ...602..257B}.  The follow-up observation with Suzaku
found that the hard emission is not a single power-law but can be
described either by a model of a broken power-law or synchrotron
radiation \citep{2009PASJ...61S.175Y}, indicating the presence of
non-thermal X-ray emission due to shock-accelerated electrons.

In our Galaxy, the X-ray study of SBs has been limited. This may be
due to their possible soft X-ray spectra, so that most of the SBs, if
any, are invisible because of the large absorption through the
Galactic plane.  Diffuse X-rays from the Cygnus super bubble (e.g.,
\cite{1980ApJ...238L..71C}), Orion-Eridanus
\citep{1995ApJ...453..256G}, Gemini-Monoceros
\citep{1996ApJ...463..224P}, and possibly the North Polar Spur (e.g.,
\cite{1995A&A...294L..25E}) are a few examples of the Galactic SB
candidates. The spectra of these sources are all soft with a plasma
temperature around 0.1--0.3~keV.  These would be near-by objects,
because of their low absorption ($N_{\rm H} \sim (1-10) \times
10^{20}$~cm$^{-2}$), their positions at high Galactic latitude, and
the apparent large ($\sim 25^{\circ} \times 25^{\circ}$) angular
sizes.  However the real distances of these sources are very
uncertain, and hence many important physical parameters such as the
physical size, the total luminosity, and the density of these SBs are
not determined well.  Furthermore these form a very limited and biased
sample of the Galactic SBs, because the majority should be prevailing
in sites of massive star formation which mainly reside in the Galactic
plane.

If a sizeable number of the SBs in our Galaxy have higher temperatures
and/or hard components, we may be able to detect these SBs by X-rays.
The Galactic center (GC) is a good region for this study since the
distance is well known.  The purpose of this paper is to search for SB
candidates in the GC and to study their physical properties.

In this paper, we briefly explain our GC observation with Suzaku
\citep{2007PASJ...59S...1M} and the data reduction in
section~\ref{section:observation_and_reduction}.  The results of the
imaging and spectral analyses are described in
section~\ref{section:analysis}. We then discuss a new SB candidate in
the GC, together with the physical properties of the bright X-ray
clump, in section~\ref{section:discussion}.  A summary is given in
section~\ref{section:summary}.  Throughout this paper, the quoted
errors represent the 90\% confidence limits, unless otherwise
mentioned.  We also adopt Galactic coordinates in which the north and
east are defined to be the positive Galactic latitude and longitude,
respectively.

\section{Observation and Data Reduction}
\label{section:observation_and_reduction}
\begin{table*}[thp]
\begin{center}
\caption{Observation log of the GC regions}
\label{table:observation_log}
\begin{tabular}{lccccc}
\hline
Target & SCI & Obs. ID 
         & Start time (UT) & End time (UT)
         & Exposure\footnotemark[$*$] \\
\hline
Sgr A East & off & 100027010
         & 2005-09-23T07:18:25 & 2005-09-24T11:05:19 & 45~ks\\
           &     & 100037040
         & 2005-09-30T07:43:01 & 2005-10-01T06:21:24 & 43~ks\\
Sgr A West & off & 100027020
             & 2005-09-24T14:17:17 & 2005-09-25T17:27:19 & 43~ks\\
           &     & 100037010
             & 2005-09-29T04:35:41 & 2005-09-30T04:29:19 & 44~ks\\
GC South & on & 501008010
         & 2006-09-26T14:18:16 & 2006-09-29T21:25:14 & 142~ks\\
GC South BGD & on & 501009010
            & 2006-09-29T21:26:07 & 2006-10-01T06:55:29 & 56~ks\\
\hline
\multicolumn{5}{@{}l@{}}{\hbox to 0pt{\parbox{180mm} {\footnotesize
 \footnotemark[$*$] Effective exposure of the screened XIS data
 }\hss}}
\end{tabular}
\end{center}
\end{table*}

The X-ray CCD cameras (X-ray Imaging Spectrometer, XIS;
\cite{2007PASJ...59S..23K}) on board Suzaku, combined with the X-Ray
Telescopes (XRTs; \cite{2007PASJ...59S...9S}), provide excellent
imaging spectroscopy capabilities.  In particular, given the low
background of the XIS and the high throughput of the XRTs, Suzaku is
suitable for detecting the diffuse X-ray emission with low surface
brightness.

Suzaku observed the Sgr A East region and its western region
(hereafter Sgr A West) twice on 2005 September, during the Performance
Verification phase \citep{2007PASJ...59S.245K}.  The southern and
northern regions, referred to as GC South and GC South BGD hereafter,
were subsequently observed on 2006 September (in the AO-1 cycle) for
the purpose of mapping the GC.  The XIS was operated with the normal
clocking and the full window modes \citep{2007PASJ...59S..23K} in all
of the observations.  Furthermore, in the last two observations, the
Spaced-row Charge Injection (SCI; \cite{2004SPIE.5501..111B},
\cite{2008PASJ...60S...1N}) was applied to improve the time-degraded
energy resolution.  The observation log is summarized in
table~\ref{table:observation_log}.

We analyzed the XIS data using the processing version of
2.0.6.13\footnote{http://www.astro.isas.jaxa.jp/suzaku/process/history/v20613.html}.
In this processing, the energy scale of the XIS data is calibrated
to within 10~eV at 6~keV, irrespective of applying the SCI
\citep{2007SPIE.6686E..22U}.  We followed the standard criteria as recommended by the Suzaku XIS team
\footnote{http://www.astro.isas.jaxa.jp/suzaku/process/v2changes/criteria\_xis.html}
for screening the XIS data.  We first filtered out the data during the
times of passage of South Atlantic Anomaly and of low cut-off rigidity ($< 6$~GV).
We then used the data taken with the Earth rim angles of $< 5^{\circ}$ and the
sun-lit Earth rim angles of $< 20^{\circ}$.  Hot and flickering pixels were
removed with \texttt{cleansis} v1.7.  Finally, we combined the events
taken in the 3$\times$3 and 5$\times$5 editing modes with
\texttt{XSELECT} v2.3.  The effective exposure times of the four regions
are $88$~ks (Sgr A East), $87$~ks (Sgr A West), $142$~ks (GC South), and
$56$~ks (GC South BGD), respectively.

Using these data sets, we searched SB candidates as described in the
following sections.  We created the XIS images and spectra with
\texttt{XSELECT} v2.3. The spectral analyses were carried out using
\texttt{XSPEC} v11.3.2.  The Redistribution Matrix File (RMF) and
Auxiliary Response File (ARF) were made via \texttt{xisrmfgen} and
\texttt{xissimarfgen} \citep{2007PASJ...59S.113I}, respectively, which
are provided through the Suzaku FTOOLS in HEASOFT v6.4.  We used the
\texttt{XIS CALDB 20080709} to construct the RMF and ARF.

\section{Analysis}
\label{section:analysis}
\begin{figure*}[hbtp]
\begin{center}
  \FigureFile(80mm,50mm){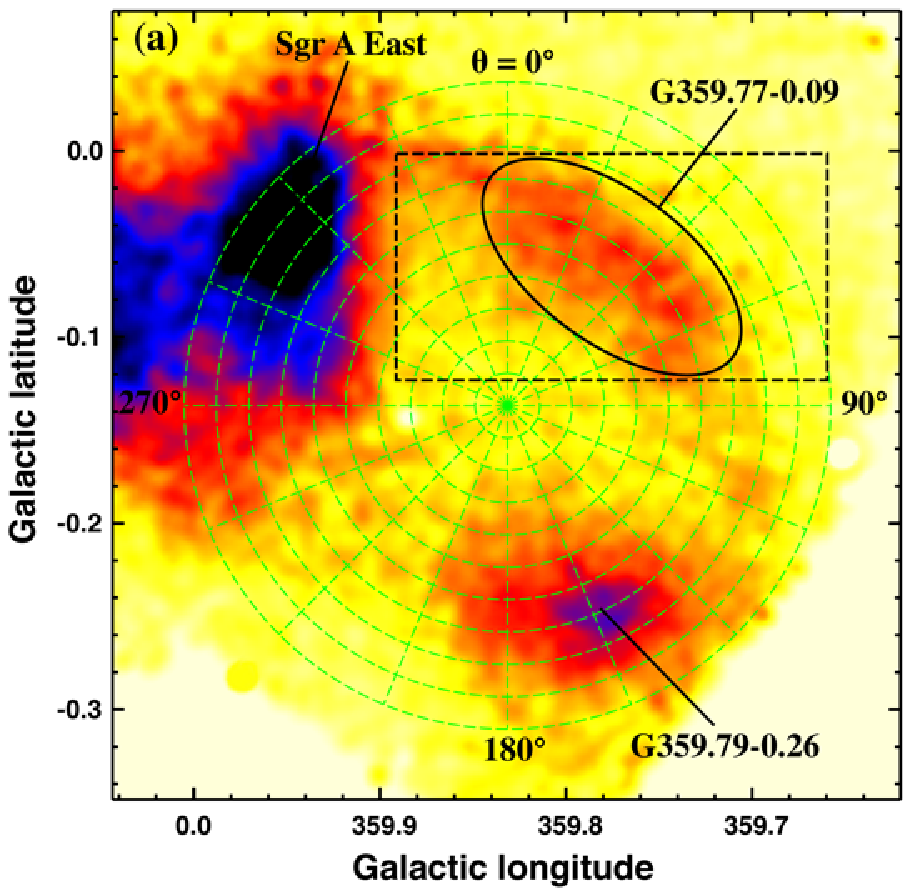}
  \FigureFile(80mm,50mm){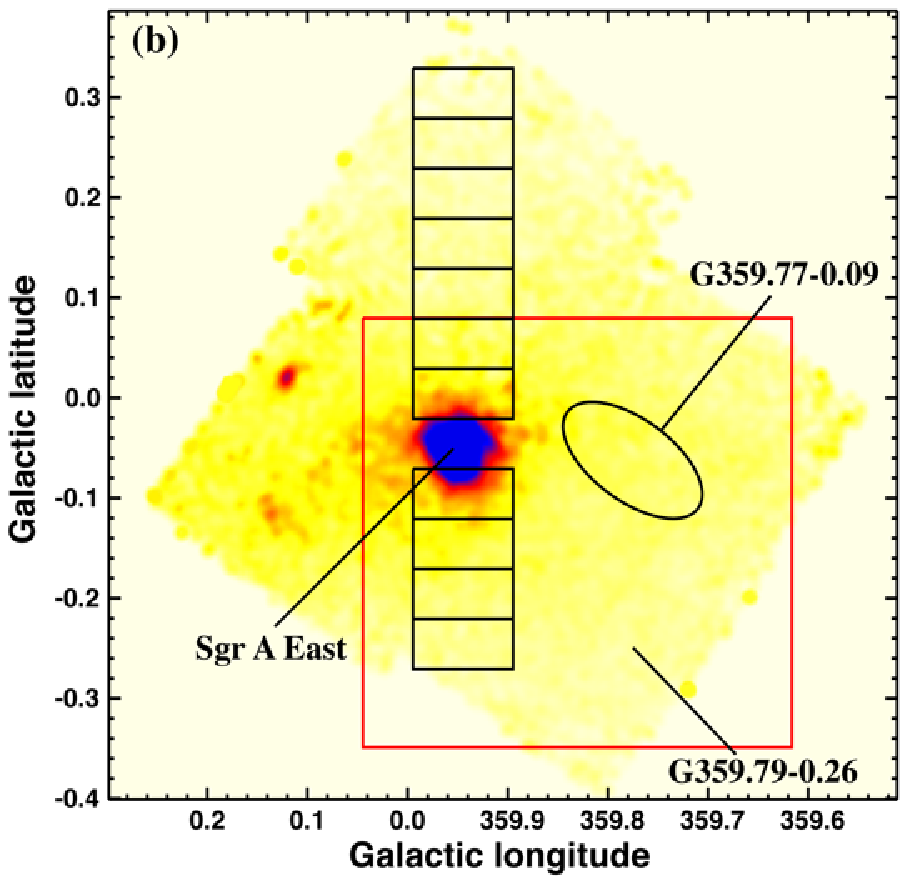}
\end{center}
\caption{XIS mosaic images of the GC region in the K$\alpha$-line
bands of He-like sulfur (a; 2.35--2.55~keV) and He-like iron (b;
6.6--6.75~keV).  Both images were binned by $8 \times 8$ pixels and
then were smoothed with $\sigma = \timeform{1.4'}$. The exposure time
and the vignetting effect of the XRTs were corrected.  The field of
view of the K$\alpha$-line band image of He-like sulfur is indicated
by a red solid square in that of He-like iron.  The regions used to
extract the G\,359.77$-$0.09 and its background spectra are drawn by a
solid ellipse and a dashed rectangle, respectively, in the
K$\alpha$-line band image of He-like sulfur.  A green dotted ``spider
web'' and a series of rectangles were used to investigate the radial
surface brightness profiles (see
section~\ref{subsection:large_scale_structure}) and the flux
distribution along the Galactic latitude (see
section~\ref{subsubsection:spatial_distribution_of_the_GCDX_emission}),
respectively.}
\label{fig:XIS_mosaic_images}
\end{figure*}

The XIS mosaic images in the K$\alpha$-line bands of He-like sulfur
(2.35--2.55~keV) and He-like iron (6.6--6.75~keV) are shown in
figure~\ref{fig:XIS_mosaic_images}.  The images extracted from the
respective observations were summed together and were binned by a factor
of 8 (1 pixel = $\timeform{1.04''} \times \timeform{1.04''}$) followed
by smoothing with the Gaussian kernel of $\sigma = \timeform{1.4'}$.  We
applied the exposure correction to the images, which includes the
vignetting effect of the XRTs.

The image of the K$\alpha$ line from He-like sulfur traces the spatial
distribution of a thermal plasma with the temperature of $\sim
10^{7}$~K.  On the other hand, that from He-like iron indicates the
spatial distribution of the Galactic Center Diffuse X-rays (GCDX;
\cite{1989Natur.339..603K}).

In the K$\alpha$-line band image of He-like sulfur
(figure~\ref{fig:XIS_mosaic_images}a), there are some bright regions
which are positionally coincident with Sgr A
East \citep{2002ApJ...570..671M}, G\,359.79$-$0.26, and G\,359.77$-$0.09;
the latter two clumps are designated by \citet{2003ANS...324..151S}.
Furthermore, we can see an elliptical ring connecting these diffuse
X-ray emission.

On the other hand, the K$\alpha$-line band image of He-like iron
(figure~\ref{fig:XIS_mosaic_images}b) shows no ring structure.  Sgr A
East is by far brightest source; its surface brightness of $> 5 \times
10^{-3}$~cts s$^{-1}$ arcmin$^{-2}$ (blue region in
figure~\ref{fig:XIS_mosaic_images}b) is an order of magnitude larger
than those of G\,359.79$-$0.26 and G\,359.77$-$0.09 ($\sim 5 \times
10^{-4}$~cts s$^{-1}$ arcmin$^{-2}$).  The surface brightness of the
inside region of the ring is the same as those of the two clumps.
Hence, the absence of the ring structure in
figure~\ref{fig:XIS_mosaic_images}b is not caused by any color
contrast effects.  Furthermore, since the ratio of the surface
brightness of Sgr A East ($\sim 4 \times 10^{-4}$~cts s$^{-1}$
arcmin$^{-2}$) to that of G\,359.79$-$0.26 ($\sim 1 \times
10^{-4}$~cts s$^{-1}$ arcmin$^{-2}$) is at most 4 in the
K$\alpha$-line band of He-like sulfur, the additive higher-temperature
plasma probably exists towards Sgr A East; \citet{2007PASJ...59S.237K}
indicated the presence of the 6~keV ($\sim 7 \times 10^{7}$~K) plasma
by the spectral analysis of Sgr A East.  Therefore it may be possible
that there is a ring structure composed of the $\sim 10^{7}$~K
temperature plasma and Sgr A East is unrelated to this object.

For the two bright regions in the ring structure of
figure~\ref{fig:XIS_mosaic_images}a, the spectral analysis of SNR
G\,359.79$-$0.26 has been already carried out by
\citet{2008PASJ...60S.183M}.  \citet{2003ANS...324..151S} reported
that G\,359.77$-$0.09 shows a thermal spectrum with the plasma
temperature of $\sim 1$~keV.  However, its origin is not clear.  We
therefore first generated and analyzed the X-ray spectrum of
G\,359.77$-$0.09 in section~\ref{subsection:G359.77-0.09}.  We
describe a more quantitative analysis of the ring structure in
section~\ref{subsection:large_scale_structure} and
\ref{subsection:ring_spectra}.

\subsection{G\,359.77$-$0.09}
\label{subsection:G359.77-0.09}
We extracted the source spectrum from the elliptical region shown in
figure~\ref{fig:XIS_mosaic_images}a.  The center of the source region
is $(l, b) = (\timeform{+359.7751D},\timeform{-0.0623D})$.  The
semi-major and semi-minor axes of the region are $\timeform{4.9'}$ and
$\timeform{2.4'}$, respectively.  We selected the periphery of the
source region as a local background region (the dashed rectangle
excluding the source region in figure~\ref{fig:XIS_mosaic_images}a).

\begin{figure}
\begin{center}
  \FigureFile(80mm,50mm){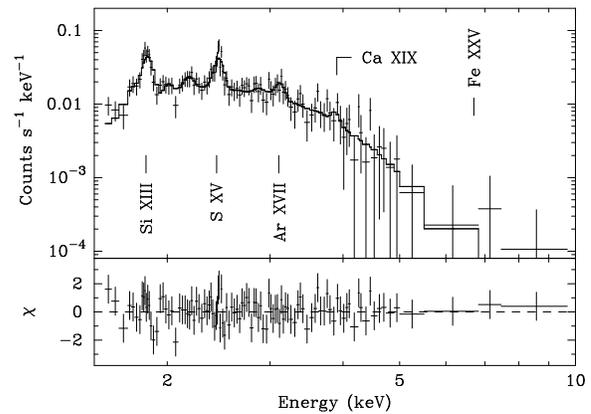}
\end{center}
\caption{Background-subtracted spectrum of G\,359.77$-$0.09. 
The solid line represents the best-fit thin thermal plasma
model.}
\label{fig:background-subtracted_spectrum}
\end{figure}

We first subtracted the non X-ray background (NXB;
\cite{2008PASJ...60S..11T}) from the source and local background
spectra and then calculated the effective areas for the source and
background regions with \texttt{xissimarfgen} in order to investigate
the difference of the energy-dependent telescope vignetting.  We then
applied the vignetting correction to the background spectra.  Finally,
the G\,359.77$-$0.09 spectrum was made by subtracting the
vignetting-corrected background spectrum.  The background-subtracted
spectrum is shown in figure~\ref{fig:background-subtracted_spectrum},
which clearly shows the K$\alpha$ lines from the He-like ions of Si,
S, and Ar, while the Ca- and Fe-K$\alpha$ lines are marginal.

We fitted the spectrum with a thin thermal plasma model in collisional
ionization equilibrium (CIE) absorbed by photoelectric absorption with
the cross sections given in \citet{1992ApJ...400..699B}.  The
elemental abundances of Si, S, and Ar were free parameters.  The
plasma temperature of $kT_{\rm e} = 0.66$~keV and the solar abundances
gave an acceptable fit ($\chi ^{2} = 73/100$~d.o.f.).  The absorption
column density is $N_{\rm H} = 6.9 \times 10^{22}$~cm$^{-2}$.  The
best-fit parameters are summarized in
table~\ref{table:best-fit_parameters_of_spectra}.  The best-fit model
is given with the solid line in
figure~\ref{fig:background-subtracted_spectrum}.  We note that we
tried to fit the spectrum with a non-equilibrium ionization plasma
model instead of the CIE plasma model.  The best-fit parameter of the
ionization timescale ($n_{e} t$) is $8.0^{+490}_{-4.6} \times
10^{11}$~cm$^{-3}$~s.  Moreover, those of the temperature ($kT_{\rm e}
= 0.72$~keV) and absorption ($N_{\rm H} = 6.3 \times
10^{22}$~cm$^{-2}$) are consistent with those obtained with the CIE
plasma model.  Thus, we adopt the best-fit parameters of the CIE
plasma in the following discussion.

\begin{table*}[htbp]
\begin{center}
\caption{Best-fit parameters of the G\,359.77$-$0.09 and the GC ring spectra\footnotemark[$*$]}
\label{table:best-fit_parameters_of_spectra}
\begin{tabular}{lccc}
\hline
Parameters & \multicolumn{3}{c}{Values} \\
           & G\,359.77$-$0.09 & GC ring & GC faint ring\\
\hline
$N_{\rm H}$ ($10^{22}$~cm$^{-2}$) & 6.9 (6.1--7.6) & 5.5 (5.2--5.7) & 5.5 (5.0--6.0) \\
$kT_{\rm e}$ (keV)                    & 0.66 (0.57--0.74) & 0.91 
 (0.88--0.94) & 0.96 ( 0.88--1.07) \\
$Z_{\rm Si}$\footnotemark[$\dagger$]   & 0.9 (0.7--1.3) & 1.0 (0.9--1.2) & 0.9 (0.6--1.2)\\
$Z_{\rm S}$\footnotemark[$\dagger$]    & 0.7 (0.6--1.0) & 1.2
 (1.1--1.3) & 1.3 (1.0--1.6) \\
$Z_{\rm Ar}$\footnotemark[$\dagger$]   & 0.9 (0.3--1.7) & 1.4 (1.1--1.7) & 1.6 (1.0--2.2) \\ 
Normalization\footnotemark[$\ddagger$]   & 0.08 (0.05--0.14) & 0.06
 (0.05--0.07) & 0.016 (0.012--0.022) \\
Flux (1.5--5~keV)\footnotemark[$\S$] & 3.0 $\times$ 10$^{-4}$ & 7.2
 $\times$ 10$^{-4}$ & 2.2 $\times$ 10$^{-4}$ \\
\hline
$\chi ^{2}$/d.o.f.             & 73/100 = 0.73 & 143/152 = 0.94 & 62/98 = 0.63 \\
\hline
\multicolumn{4}{@{}l@{}}{\hbox to 0pt{\parbox{160mm} {\footnotesize
 \footnotemark[$*$] The values in parentheses represent the 90\% confidence intervals.
 \par\noindent
 \footnotemark[$\dagger$] Elemental abundances relative to solar. 
 \par\noindent
 \footnotemark[$\ddagger$] In unit of $10^{-14}/(4 \pi D^{2}) \int
                           n_{\rm e} n_{\rm H} dV$~cm$^{-5}$. Here $V$
                           and $D$ are the volume and distance to the
                           plasma, respectively.
 \par\noindent
 \footnotemark[$\S$] Absorbed flux in units of photons s$^{-1}$~cm$^{-2}$.
}\hss}}
\end{tabular}
\end{center}
\end{table*}

\subsubsection{Spatial distribution of the GCDX emission}
\label{subsubsection:spatial_distribution_of_the_GCDX_emission}
The GCDXs are the most serious source of local background for
G\,359.77$-$0.09, because the GCDX flux is position-dependent
\citep{2007PASJ...59S.245K} and G\,359.77$-$0.09 is a faint diffuse
source located near the GC on the Galactic plane ($b =
\timeform{-0.05D}$).  We therefore checked whether or not the GCDX
subtraction wad done properly.

The K$\alpha$ line from He-like iron (6.7~keV) is a good indicator of
the GCDX emission (see figure~\ref{fig:XIS_mosaic_images}b).  The
6.7~keV line fluxes along the Galactic longitude are available from
\citet{2007PASJ...59S.245K}, so that we examined the 6.7~keV line
fluxes along the Galactic latitude.

We extracted spectra in the 5.5--11.5~keV band from a series of $6'
\times 3'$ rectangular regions in figure~\ref{fig:XIS_mosaic_images}b.
We fixed the central longitude of each rectangle to be $l =
\timeform{+359.9443D}$, the same Galactic longitude as that of Sgr
A$^{*}$.  After the NXB subtraction, each spectrum was fitted by the
same phenomenological model as that used in \citet{2007PASJ...59S.245K};
an absorbed thermal bremsstrahlung with several Gaussian emission lines.

\begin{figure}
\begin{center}
  \FigureFile(80mm,50mm){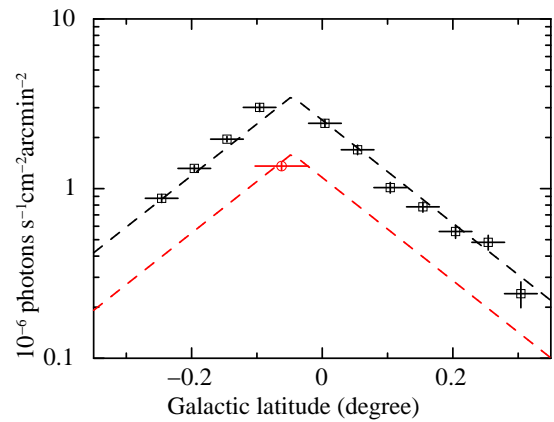}
\end{center}
\caption{The 6.7~keV line fluxes of G\,359.77$-$0.09 (red circle) and
 the ladder area along the Galactic latitude (black squares).  The black
 and red dotted lines represent the best-fit models of the flux
 distribution along the Galactic latitude at $l = \timeform{+359.9443D}$
 and $l = \timeform{+359.7751D}$, respectively.
}
\label{fig:FeXXV_distribution}
\end{figure}

We plot the best-fit parameters of the 6.7~keV line fluxes along the
Galactic latitude in figure~\ref{fig:FeXXV_distribution}.  Although the
flux distribution is highly asymmetric along the Galactic longitude with
respect to the Galactic center \citep{2007PASJ...59S.245K}, that along
the Galactic latitude at $l = \timeform{+359.9443D}$ is roughly
symmetric with respect to the Galactic plane.  The 6.7~keV line fluxes
are decreased exponentially with the distance from the Sgr A$^{*}$,
except for the direction along the Galactic positive longitude.  We
therefore fitted the 6.7~keV line fluxes along the Galactic negative
longitude and latitude (both positive and negative latitudes)
simultaneously with an exponential model given by,
\begin{equation}
\label{eq:spatial_dist_of_6.7keV_line_flux} F(l, b) = A \exp \Bigl( -
\frac{|l - l_{c}|}{\sigma _{l}} - \frac{|b - b_{c}|}{\sigma _{b}}
\Bigr),
\label{equation:expnential_model}
\end{equation}
where $l_{c}$ and $b_{c}$ are fixed to the position of Sgr A$^{*}$.  We
obtained the best-fit parameters as follows; $A = (3.5 \pm 0.2) \times
10^{-6}$~photons s$^{-1}$~cm$^{-2}$~arcmin$^{-2}$, $\sigma _{l} =
\timeform{0.22D} \pm \timeform{0.02D}$, and $\sigma _{b} =
\timeform{0.14D} \pm \timeform{0.01D}$.  We show the best-fit model at
$l = \timeform{+359.9443D}$ with a black dashed line in
figure~\ref{fig:FeXXV_distribution}.

Assuming the validity of equation~\ref{equation:expnential_model} in the
GC region, we inferred the 6.7~keV flux distribution at the
G\,359.77$-$0.09 position ($l = \timeform{+359.7751D}$), as is indicated
by a red dashed line in figure~\ref{fig:FeXXV_distribution}.  The
estimated flux at G\,359.77$-$0.09 ($l = \timeform{+359.7751D}$, $b =
\timeform{-0.0623D}$) is $1.43 \times
10^{-6}$~photons~s$^{-1}$~cm$^{-2}$~arcmin$^{-2}$.

On the other hand, we fitted the NXB-subtracted source spectrum (X-ray
background unsubtracted G\,359.77$-$0.09 spectrum) with an absorbed
thermal bremsstrahlung plus several Gaussian emission lines.  We
obtained the best-fit 6.7~keV line flux to be $(1.36 \pm 0.07) \times
10^{-6}$~photons~s$^{-1}$~cm$^{-2}$~arcmin$^{-2}$, as is indicated by a
red circle in figure~\ref{fig:FeXXV_distribution}, and it is consistent
with the model estimation.  This consistency indicates that the 6.7~keV
line flux in the source region can be attributed to the GCDX emission.
Thus no flux at the 6.7~keV line in the G\,359.77$-$0.09 spectrum (see
figure \ref{fig:background-subtracted_spectrum}) supports that our GCDX
subtraction was done accurately.

\subsection{Large-scale ring structure}
\label{subsection:large_scale_structure}
To verify the presence of the ring structure quantitatively, we
constructed the radial surface brightness profile from the
K$\alpha$-line band image of He-like sulfur.  We divided a circular
region centered at ($l$, $b$) = ($\timeform{+359.8312D}$,
$\timeform{-0.1367D}$) into 10 annuli, each of which has a radial width
of $\timeform{1'}$, and separated the respective annuli into 16 sectors
with an azimuthal angle of $\timeform{22.5D}$.  We show these sectors
with a green dotted spider web in figure~\ref{fig:XIS_mosaic_images}a.

\begin{figure}[htp]
\begin{center}
\FigureFile(80mm,50mm){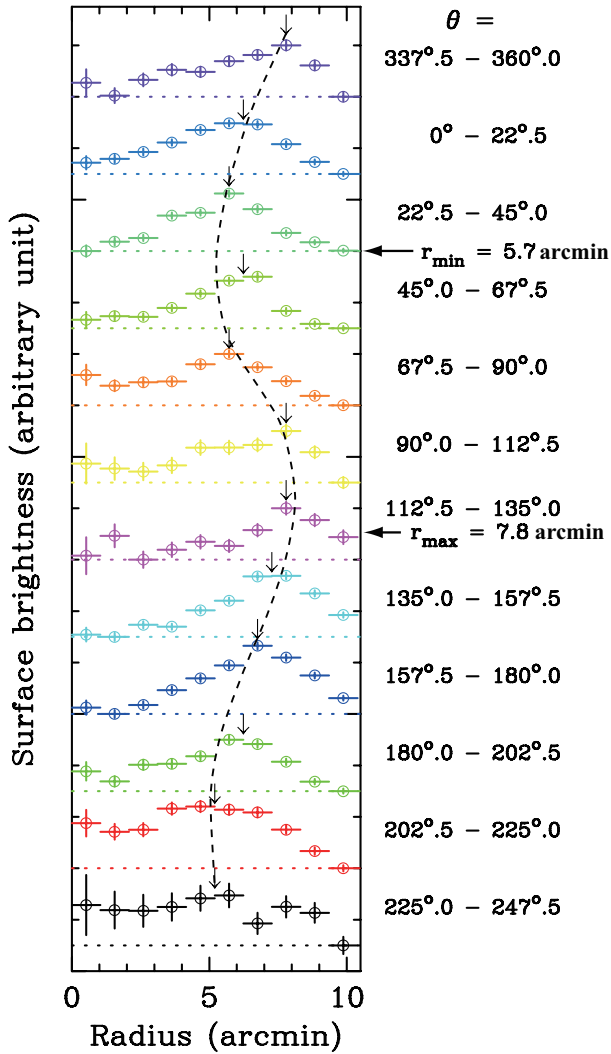}
\end{center}
\caption{Radial surface brightness profiles in the 12 azimuthal
directions.  The error bars represent the 1$\sigma$ confidence limits.
For clarity, the vertical axis of each profile is scaled arbitrarily.
The background level of each profile is represented by a dotted line.
The profiles are shown in clockwise order from the north-northeast
direction (top) to the east-east south direction (bottom).  The
azimuthal angle of each profile (north is $\theta = 0^{\circ}$ and west
is $\theta = 90^{\circ}$; see figure~\ref{fig:XIS_mosaic_images}a) is
given in the right side of the panel.  To guide the eye, we show a
sinusoidal line (dashed line) approximately following the brightness peak
in each direction (indicated by arrows).}
\label{fig:sb_radial_profile}
\end{figure}

The radial profiles in the 12 directions are shown in
figure~\ref{fig:sb_radial_profile}.  The uppermost profile corresponds
to the north-northeast direction ($\theta =
\timeform{337.5D}$--$\timeform{0.0D}$ in
figure~\ref{fig:XIS_mosaic_images}a), and the profiles in the other
directions are placed from top to bottom in the clockwise order.  We
ignored the 4 sectors from the east-eastsouth to the north-northeast,
where the contamination from Sgr A East is significant because of the
$\sim 2'$ angular resolution of the XRTs.

We connect the brightness peak of each profile (indicated by arrows)
with a smooth dashed line.  Since the peak positions are in the range of
$\timeform{5.7'}$--$\timeform{7.8'}$ and comprise an S-shaped line, we
suggest that there is a large elliptical ring toward the GC.  Since each
profile has a width of $6'$--$9'$ around the peak, the size of the outer
boundary of the ring is estimated to be $\sim 20' \times 16'$.

\begin{figure*}
\begin{center}
  \FigureFile(80mm,50mm){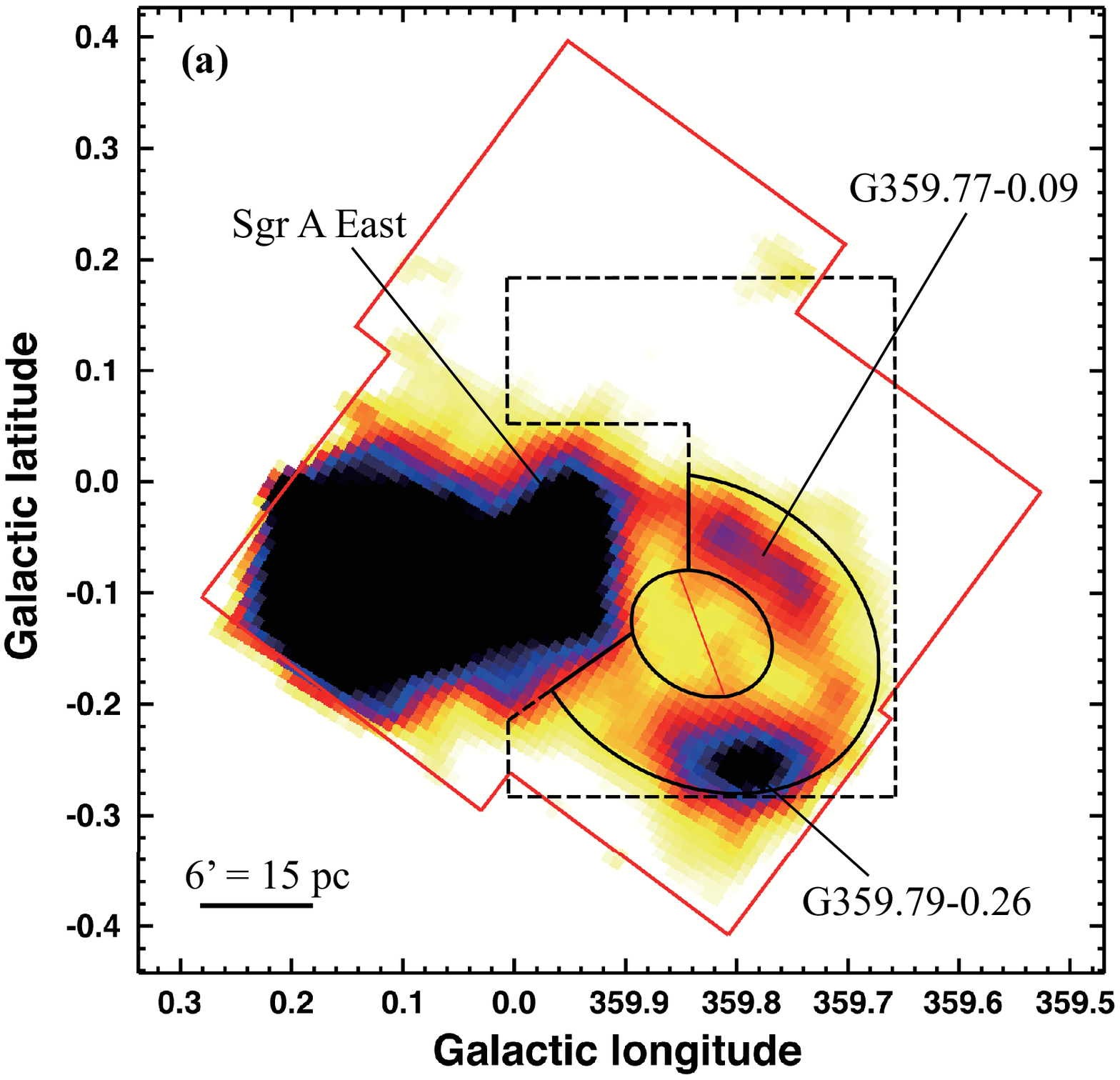}
  \FigureFile(80mm,50mm){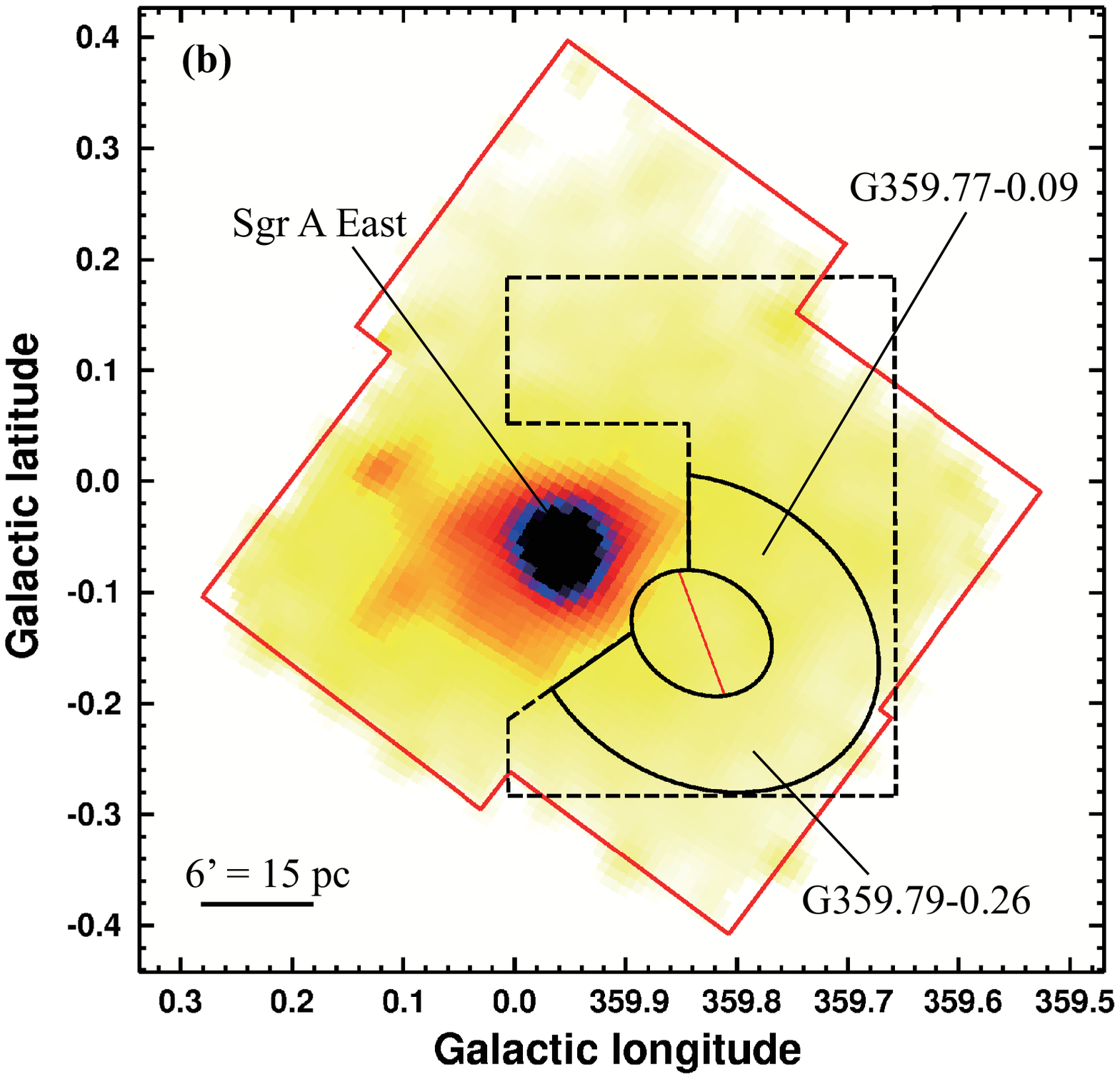}
\end{center}
\caption{GC mosaic images of He-like sulfur K$\alpha$-line (a), and
 He-like and H-like iron K$\alpha$-lines (b).  The images were binned by
 $32 \times 32$ pixels and the continuum fluxes were then subtracted (see
 text) and the images were smoothed with $\sigma =
 \timeform{1.66'}$. The exposure time and the vignetting effect of the
 XRTs were corrected.  The XIS field of view is indicated by a red
 polygon.  The GC-ring spectrum is extracted from the elliptical
 arc-like region encircled with solid curves.  The background region is
 shown by a dashed polygon excluding the source-extraction region and an
 ellipse with a red slash. }
\label{fig:emission_line_images}
\end{figure*}

We next made a GC image consisting of the K$\alpha$ emission from
He-like sulfur.  We created the narrow-band images in 2.39--2.52
(hereafter image A), 1.9--2.3 (image B), and 2.6--3.0~keV (image C) for
each observation.  The images in the latter two energy bands are used to
estimate the continuum component in the 2.39--2.52~keV band.  These
images were binned by $32$~pixel $\times$ $32$~pixel ($\timeform{0.55'}
\times \timeform{0.55'}$) and summed to make the GC mosaic images.
After the scaling of the image-extraction energy ranges, we subtracted
the image B and C from the image A.  The continuum-subtracted GC image
was corrected for the exposure and vignetting effects and then smoothed
with the Gaussian of $\sigma = \timeform{1.66'}$.  The K$\alpha$ line
image of He-like sulfur is shown in
figure~\ref{fig:emission_line_images}a.

The image again shows a ring-like structure; the surface brightness of
the connecting region between G\,359.77$-$0.09 and G\,359.26$-$0.79 and
that between G\,359.26$-$0.79 and Sgr A East is $\sim 2$ times larger
than that of the inside region of the ring.  Furthermore, we
investigated the significance of detecting the K$\alpha$ emission from
He-like sulfur.  Assuming that each $32$~pixel $\times$ $32$~pixel
($\timeform{0.55'} \times \timeform{0.55'}$) area contains X-ray photons
of $C_{\rm A}$ for the image A, $C_{\rm B}$ for the image B, and $C_{\rm
C}$ for the image C, the significance level (S/N ratio) is calculated by
$(C_{\rm A} - C_{\rm B}f_{\rm B} - C_{\rm C}f_{\rm C}) / \sqrt{C_{\rm A}
+ C_{\rm B}f^{2}_{\rm B} + C_{\rm C}f^{2}_{\rm C}}$, where $f_{\rm B}$
and $f_{\rm C}$ are the scaling factors due to the image-extraction
energy ranges.  The S/N ratio inside the ring (indicated by yellow color
in figure~\ref{fig:emission_line_images}a) region is $1.5$--$2 \sigma$
in each unit area ($32$~pixel $\times$ $32$~pixel), while that on the
ring region (indicated by orange color) is about $2.5$--$3 \sigma$.

We also made a GC image due to the K$\alpha$ emission from He-like and
H-like iron (see figure~\ref{fig:emission_line_images}b).  Here we
selected the energy bands of 6.5--7.0~keV (image A), 4.5--6.0~keV (image
B), and 7.2--8.0~keV (image C).  The image was made by the same method
as that of the K$\alpha$ emission from He-like sulfur.  As is shown in
figure~\ref{fig:XIS_mosaic_images}b, we can see that there is the strong
emission centered at Sgr A East and that the ring structure in
figure~\ref{fig:emission_line_images}a is absent.

\subsection{The Ring Spectra}
\label{subsection:ring_spectra}
We extracted a spectrum from the elliptical arc-like region encircled
with solid curves in figure~\ref{fig:emission_line_images}a.  We removed
a part of the ring structure (hereafter GC ring) where the contamination
from Sgr A East cannot be negligible from the source-extraction region.
The sizes of the outer and inner elliptical arcs are $20' \times 16'$
and $\timeform{8.0'} \times \timeform{6.4'}$, respectively.  The
background region is drawn by a dashed polygon excluding the
source-extraction region. We also added the inside of the GC ring
(indicated by an ellipse with a red slash) to the background region.

\begin{figure*}[hbtp]
\begin{center}
  \FigureFile(80mm,50mm){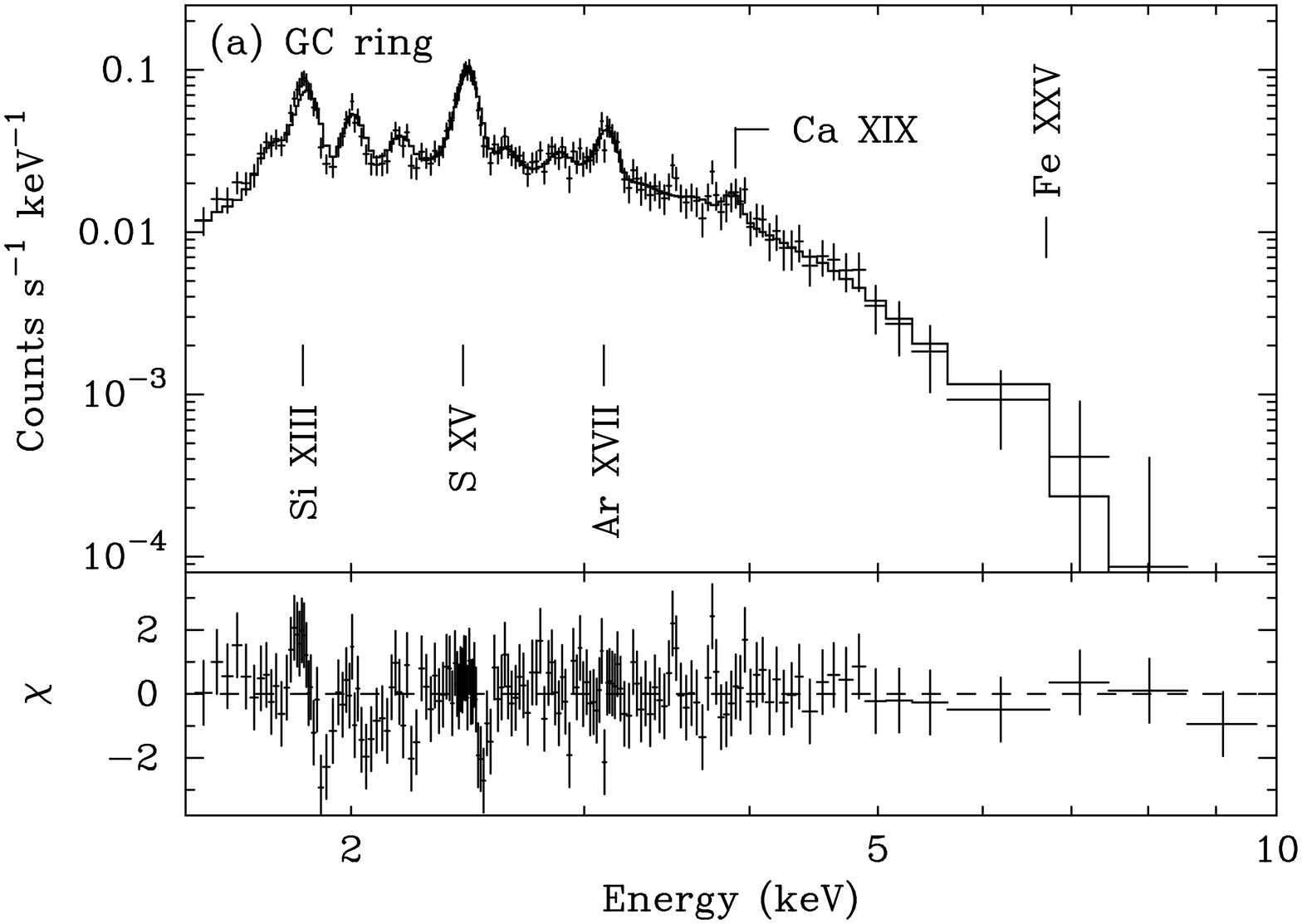}
  \FigureFile(80mm,50mm){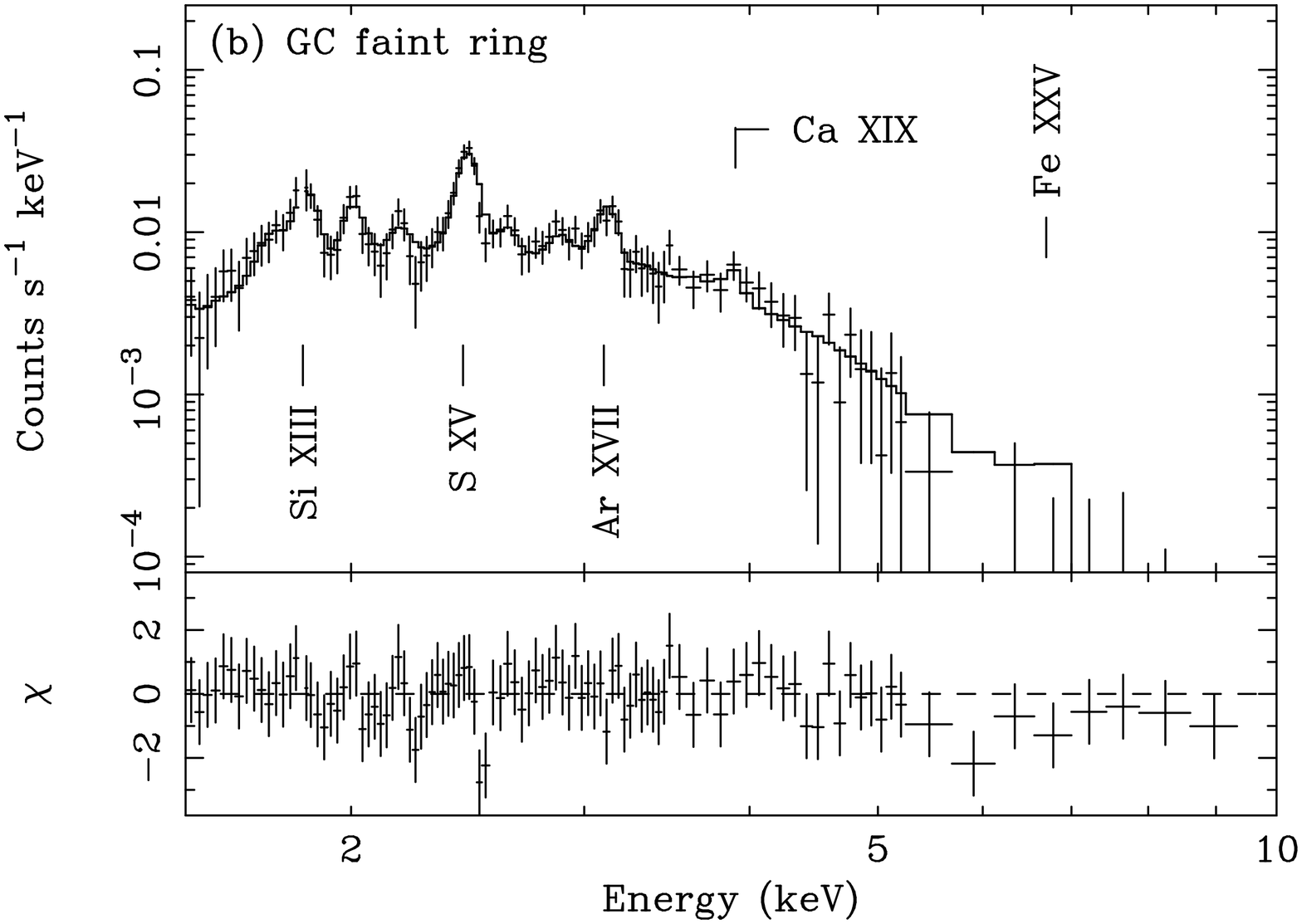}
\end{center}
\caption{Background-subtracted spectra of the GC ring (a) and the GC
faint ring (b).  The solid line represents the best-fit thin thermal
plasma model.}
\label{fig:background-subtracted_GCring_spectrum}
\end{figure*}

The GC-ring spectra were extracted from the GC South and Sgr A West
observations, while the background spectra are done from the GC South,
GC South BGD, and Sgr A West observations.  We created the
corresponding NXB spectra by \texttt{xisnxbgen} which were subtracted
from the ring and background spectra.  For all the spectra the
energy-dependent vignetting corrections described in
section~\ref{subsection:G359.77-0.09} were applied.  The ring and
background spectra were normalized by the effective area and were
summed together.  Finally, we subtracted the background spectrum from
the ring spectrum, which is shown in
figure~\ref{fig:background-subtracted_GCring_spectrum}a.

The spectrum clearly shows the Si, S, and Ar emission lines, in addition
to a marginal emission line due to He-like Ca.  We fitted the
spectrum by an absorbed CIE plasma model with the variable abundances
of Si, S, and Ar.  The best-fit parameters of the absorption column
density and the electron temperature are $N_{\rm H} = 5.5 \times
10^{22}$~cm$^{-2}$ and $kT_{\rm e} = 0.91$~keV.  The elemental
abundances of Si, S, and Ar are solar or subsolar values.  The results
are summarized in table~\ref{table:best-fit_parameters_of_spectra}.  We
note that all the best-fit parameters of the absorption column density,
the electron temperature, and the elemental abundances are intermediate
values between those of G\,359.77$-$0.09 (see section
\ref{subsection:G359.77-0.09}) and G\,359.79$-$0.26 (see table 3 of
\cite{2008PASJ...60S.183M}).

Furthermore, we made a ring spectrum excluding the G\,359.77$-$0.09
and G\,359.79$-$0.26 emission to examine whether the region with the
faint K$\alpha$-line emission from He-like sulfur (referred as to
faint ring hereafter) is related to these two bright clumps.  The
excluded regions for G\,359.77$-$0.09 and G\,359.79$-$0.26 are the
same as that described in section~\ref{subsection:G359.77-0.09} (see
also the solid ellipse in figure~\ref{fig:XIS_mosaic_images}a) and that
described in section 3 of \citet{2008PASJ...60S.183M}, respectively.
As is shown in figure~\ref{fig:emission_line_images}b, the periphery
of G\,359.77$-$0.09 shows relatively bright K$\alpha$-line
emission from highly ionized iron because of its location.  Thus, in
order to extract the corresponding background spectrum, we removed the
area overlapped with the dashed rectangle shown in
figure~\ref{fig:XIS_mosaic_images}a from the background region.  We
again subtracted the background spectrum after applying the vignetting
correction.  The X-ray spectrum of the faint ring is shown in
figure~\ref{fig:background-subtracted_GCring_spectrum}b.

The spectrum shows the K$\alpha$ emission lines from He-like Si, S,
and Ar.  This feature is similar to that of the ring spectrum shown in
figure~\ref{fig:background-subtracted_GCring_spectrum}a, and indicates
that the X-ray emission has a thin thermal plasma origin.  We fitted
the faint ring spectrum with an absorbed CIE plasma model.  The
elemental abundances of Si, S, and Ar were allowed to vary again.  The
best-fit parameters are consistent with those derived from the ring
spectrum including the G\,359.77$-$0.09 and G\,359.79$-$0.26 emission;
the X-ray emission from the plasma with the temperature of $kT_{\rm e}
= 0.96$~keV is attenuated with the absorption of $N_{\rm H} = 5.5
\times 10^{22}$~cm$^{-2}$.  We summarized the best-fit parameters in
table~\ref{table:best-fit_parameters_of_spectra}.  This result
strengthens the physical connection of the faint ring emission to
G\,359.77$-$0.09 and G\,359.79$-$0.26.

\section{Discussion}
\label{section:discussion}
\subsection{G\,359.77$-$0.09}
\label{subsection:single_plasma}
The X-ray spectrum of G\,359.77$-$0.09 clearly shows the presence of a
thin thermal plasma ($kT_{\rm e} \sim 0.7$~keV).  The heavy absorption
of $N_{\rm H} = 6.9 \times 10^{22}$~cm$^{-2}$ indicates that the plasma
is located in the GC.  Using the best-fit plasma parameters derived from
the G\,359.77$-$0.09 spectral analysis, we can estimate some physical
properties.  Assuming that the distance to the plasma is $8.5$~kpc, the
emission measure is estimated to be $6.9 \times 10^{58}$~cm$^{-3}$.  If
the plasma is an ellipsoid with the dimensions of $\timeform{4.9'} \times
\timeform{2.4'} \times \timeform{2.4'}$, corresponding to $12$~pc
$\times$ $6.0$~pc $\times$ $6.0$~pc in the GC, the volume of the X-ray
emitting plasma is $3.9 \times 10^{58}$~cm$^{3}$.  The electron
density of the plasma is then derived to be $1.3 f^{-1/2}$~cm$^{-3}$,
where $f$ is a filling factor.

The elemental abundances of the plasma are consistent with the solar
values, which implies that the X-ray emitting plasma has an interstellar
medium (ISM) origin.  The total mass of the plasma is $M = 1.4 n_{\rm e}
m_{\rm p} V = 58 f^{1/2} M_{\odot}$ ($m_{\rm p}$ represents the proton
mass), and its thermal energy is estimated to be $E_{\rm th} =
\frac{3}{2} \frac{MkT_{\rm e}}{\mu m_{\rm p}} = 1.9 f^{1/2} \times
10^{50}$~erg.  Here $\mu$ represents the mean atomic weight of $0.604$.
The plasma temperature gives the sound speed of $v_{\rm s} = \bigl (
kT_{\rm e} \gamma / \mu m_{\rm p} \bigr ) ^{1/2} \simeq
430$~km~s$^{-1}$, when assuming the energy equipartition of electrons
and ions.  $\gamma$ represents the specific heat ratio of $5/3$.  By 
dividing the semi-major axis of the plasma ($\timeform{4.9'} \sim
12$~pc) by its velocity, we can estimate the dynamical age of the plasma
to be $t_{\rm dyn} = 2.7 \times 10^{4}$~yr.

The thermal energies of G\,359.77$-$0.09 and G\,359.79$-$0.26 are
similar to each other ($1.9 \times 10^{50}$~erg and $1.7 \times
10^{50}$~erg, respectively), although the apparent X-ray flux of the
former is 1.7 times smaller than that of the latter.  This difference
comes from the different plasma densities.  Therefore the X-ray fainter ring
regions other than the two spots may have comparable thermal energies.

\subsection{Super bubble}
\label{subsection:super_bubble}
The radial profile indicates that the brightness peaks comprise an
elliptical ring at the center of ($l$, $b$) = ($\timeform{+359.8312D}$,
$\timeform{-0.1367D}$).  The X-ray spectrum of G\,359.77$-$0.09, which
is on the elliptical annular structure, is similar to that of
G\,359.79$-$0.26, another bright spot on the elliptical annulus, while
the absorption of $N_{\rm H} = 6.9 \times 10^{22}$~cm$^{-2}$ is 1.5
times larger than that of G\,359.79$-$0.26 ($N_{\rm H} = 4.5 \times
10^{22}$~cm$^{-2}$).

\begin{figure}[htp]
\begin{center}
  \FigureFile(80mm,50mm){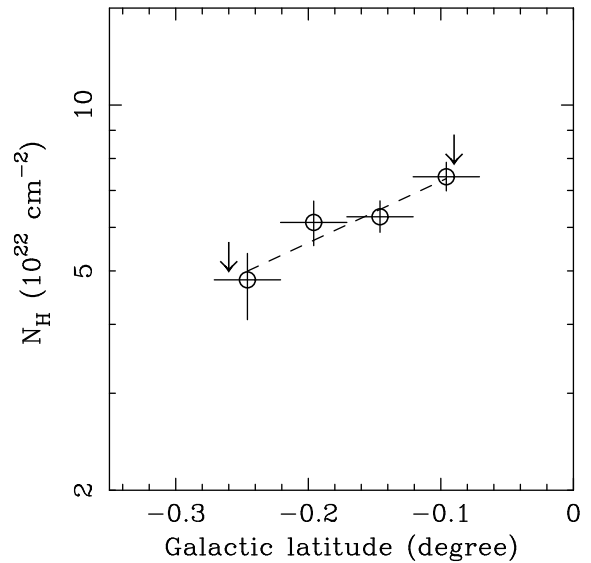}
\end{center}
\caption{Best-fit $N_{\rm H}$ parameters in the ladder region of
 figure~\ref{fig:XIS_mosaic_images}b in the Galactic negative latitude.
 The dashed line is drawn to guide the eye and the arrows show the latitude
 positions of G\,359.77$-$0.09 and G\,359.79$-$0.26.}
\label{fig:Nh-latitude}
\end{figure}

We therefore examined the $N_{\rm H}$ dependence on the Galactic
latitude near to G\,359.77$-$0.09 and G\,359.79$-$0.26 using the GCDX
spectra in the ladder of figure~\ref{fig:XIS_mosaic_images}b, because
this ladder region is free from the two sources.  The best-fit column
densities ($N_{\rm H}$) are shown in figure~\ref{fig:Nh-latitude}, where
the arrows show the latitudes of the two sources.

Using figure~\ref{fig:Nh-latitude}, the $N_{\rm H}$ ratio at the center
latitudes of the two sources is estimated to be $7.6/5.0 = 1.5$, exactly
the same as that calculated by the observed $N_{\rm H}$ ($6.9/4.5 =
1.5$).  Thus we can infer that both G\,359.77$-$0.09 and
G\,359.79$-$0.26 are located at the same distance, the GC region, and
that these two bright spots are physically associated.

The elliptical annular structure has an apparent size of $20' \times
16'$, which corresponds to $49$~pc $\times$ $40$~pc at the distance of
$8.5$~kpc.  While the spatial extent is half of a typical size of the
SBs ($\sim 100$~pc, for instance, the physical size of the
Orion-Eridanus SB is $80$~pc $\times$ $160$~pc given its distance of
$226$~pc; see \cite{1995ApJ...453..256G}), it is 4--5 times larger than
that of the nearest GC SNR, Sgr A East ($\sim 10$~pc;
\cite{2002ApJ...570..671M}).

We generated a GC-ring spectrum which averages the spectra of
G\,359.77$-$0.09, G\,359.79$-$0.26, and the connecting region, and then
derived a heavy absorption column of $N_{\rm H} = 5.5 \times
10^{22}$~cm$^{-2}$ and a plasma temperature of $kT_{\rm e} =
0.91$~keV.  Owing to the good photon statistics, we can determine
the emission measure precisely, which is calculated to be $5.2 \times
10^{58}$~cm$^{-3}$ at the GC distance.  The volume of the GC ring (i.e.,
the elliptical arc-like region in figure~\ref{fig:emission_line_images})
is estimated to be $7.6 \times 10^{59}$~cm$^{3}$ if the dimension along
the line of sight is equal to that of the semi-minor axis.  The 
electron density and mass of the plasma are then found to be $0.26 f^{-1/2}$~cm$^{-3}$ 
and $232 f^{1/2} M_{\odot}$, respectively, where $f$ again represents a
filling factor.  Consequently, the thermal energy is estimated to be
$E_{\rm th} = 1.0 f^{1/2} \times 10^{51}$~erg, consistent with the
theoretically predicted energy released by a single SN, assuming that
the filling factor is unity.

However, the thermal energies of X-ray emitting SNRs have never
reached these values, instead being typically $\sim 10^{50}$~erg, as is the case of
G\,359.77$-$0.09 and G\,359.79$-$0.26.  This inconsistency would be
caused by the partial energy transfer from a single supernova explosion
to X-rays.  The thermal energy of G\,359.77$-$0.09 and G\,359.79$-$0.26
amounts to $3.6 \times 10^{50}$~erg, if both plasmas have a filling
factor of unity.  Assuming that the energy contribution of the ring
region except for G\,359.77$-$0.09 and G\,359.79$-$0.26 is negligible,
the lower limit of the filling factor of the GC ring is estimated to be
$f \sim 0.6$.  If the filling factor is relatively large, i.e., $f >
0.6$, an additional energy source would be needed.  Therefore, we can
regard this elliptical arc-like structure as a super bubble, or no less
than 2 SNe events.

This putative GC SB has a higher temperature ($\sim 0.9$~keV) than those
of the other SBs.  We summarize the physical properties of the known
Galactic SB candidates in table~\ref{table:SB_properties}.  All the
Galactic SB candidates, except for the GC SB, show thermal emission with
a temperature of $\leq 0.3$~keV.  For the extragalactic SBs,
\citet{2004ApJ...605..751C} reported that the diffuse X-ray emission
from DEM L192, a SB hosting two OB associations in the LMC, is dominated
by a thermal component with the temperature of $\sim 0.2$~keV.  Although
30 Doradus, another SB in the LMC, shows a hard tail, the integrated
diffuse emission is fitted by a single-temperature plasma of $kT_{\rm e}
\sim 0.3$~keV (\cite{2001A&A...365L.202D}; \cite{2006AJ....131.2140T}).
Thus the high plasma temperature of the GC SB is unusual.

\begin{table*}[htp]
\begin{center}
\caption{Properties of the Galactic SB candidates}
\label{table:SB_properties}
\begin{tabular}{lccccc}
\hline
Name &
Distance (pc) &
Spatial size &
Temperature (keV) &
Density (cm$^{-3}$)&
References\footnotemark[$\ddagger$] \\
\hline
Orion-Eridanus SB & 226 & $\timeform{20D} \times \timeform{35D}$ &
0.09--0.15 & 0.015 & 1 \\
Cygnus SB & 2000 & $\timeform{18D} \times \timeform{13D}$ &
0.14--0.21 & 0.02 & 2,3 \\
Gemini-Monoceros SB & 100--1300 & $\timeform{25D} \times \timeform{25D}$ &
0.08--0.19 & 0.02--0.05\footnotemark[$*$] & 4 \\
North Polar Spur & 100--210 & $\timeform{34D} \times \timeform{34D}$\footnotemark[$\dagger$] &
0.17--0.30 & $\sim 0.03$ & 5,6,7 \\
GC SB     & 8500 & $\timeform{20'} \times \timeform{16'}$ &
0.91 & 0.26 & our work \\
\hline
\multicolumn{6}{@{}l@{}}{\hbox to 0pt{\parbox{160mm} {\footnotesize
 \footnotemark[$*$] The distance is assumed to be 300~pc.
 \par\noindent
 \footnotemark[$\dagger$] A spherical shocked plasma with a radius of
 140~pc is assumed to be at a distance of 210~pc.  The apparent size is
 ($\Delta l$, $\Delta b$) $\sim$ ($15^{\circ}$, $40^{\circ}$), based on
 the ROSAT 3/4~keV map.
 \par\noindent
 \footnotemark[$\ddagger$] (1) \citet{1995ApJ...453..256G}, (2)
 \citet{1980ApJ...238L..71C}, (3) \citet{2001A&A...371..675U}, (4)
 \citet{1996ApJ...463..224P}, (5) \citet{1995A&A...294L..25E}, (6)
 \citet{2003MNRAS.343..995W}, (7) \citet{2008PASJ...60S..95D}
}\hss}}
\end{tabular}
\end{center}
\end{table*}

Furthermore, the plasma density of the GC SB is at least 10 times larger
than those of the other SB candidates.  Whether the high temperature and
the large density are related to some environmental factors peculiar to
the GC is an open question.  The high temperature of the plasma may also
imply hidden non-thermal components or something else.

Some Galactic SBs are known to be related to OB associations.
Orion-Eridanus SB is possibly originated from the Orion OB1 association
and a following SN explosion with an age of $\sim 1.2 \times
10^{5}$~yr \citep{1995ApJ...453..256G}.  Cygnus SB and the North Polar Spur
may obtain their thermal energy from the Cyg OB2 association
\citep{2001A&A...371..675U} and the Sco-Cen OB association
\citep{1995A&A...294L..25E}, respectively.  Thus, we investigated the
presence of the OB associations in the GC, using two catalogs of X-ray
point sources towards the GC, the Chandra Galactic Center Point Source
Catalog (abbreviated to changalcen; \cite{2003ApJ...589..225M}) and the
Chandra Galactic Central 150 Parsecs Source Catalog (abbreviated to
chanc150pc; \cite{2006ApJS..165..173M}).  We applied both hardness and
flux selection to the sources, and found that $\sim 50$ sources are
located within the GC SB.  However, all the sources are not brighter
than $10^{34}$ erg s$^{-1}$ in the 0.5--8.0~keV band, when we assume the
spectral model and the source distance to be an absorbed power-law ($N_{\rm
H} = 6.0 \times 10^{22}$~cm$^{-2}$ and $\Gamma = 1.7$) and a distance of 8.5~kpc.
This examination is consistent with the fact that no candidate of the OB
association has been reported in the GC.  Hence, it is plausible that
the GC SB is formed by multiple (no less than 2) SN explosions.

For the future, a deep X-ray survey in the GC would be required not
only to detect the OB association but also to search other SB
candidates which may be hidden in the GC.

\section{Summary}
\label{section:summary}
We summarize the results derived from the G\,359.77$-$0.09 and GC-ring analyses
below:
\begin{enumerate}
\item The X-ray spectrum of G\,359.77$-$0.09 is well fitted by a thin
      thermal plasma with a temperature of $\sim 0.7$~keV.  The large
      absorption towards the plasma ($N_{\rm H} = 6.9 \times
      10^{22}$~cm$^{-2}$) indicates that the plasma is located in the
      GC.  Since the elemental abundances of Si, S, and Ar are
      consistent with the solar values, the X-ray emitting plasma is
      possibly originated from a shock-heated ISM.
\item The radial surface brightness profile constructed from the
      K$\alpha$-line band image of He-like sulfur indicates the presence
      of a large ($20' \times 16'$) ring-like object, which connects two
      bright X-ray clumps, G\,359.77$-$0.09 and G\,359.76$-$0.26.  The
      absorption column densities of these clumps show the $N_{\rm H}$
      dependence on the Galactic latitude near the GC.  Furthermore, the
      X-ray spectrum of the ring-like structure shows the presence of a
      thin thermal plasma. The thermal energy of the plasma is estimated
      to be $1.0 \times 10^{51}$~erg, which exceeds that released by a
      single X-ray detected SNR.  Hence, we propose that the ring-like
      structure is a super bubble candidate in the GC.
\end{enumerate}

\bigskip
We are grateful to an anonymous referee for his/her fruitful comments
and suggestions to improve our paper.  We also would like to thank all
the Suzaku team members for their support of the observation and
useful information on the XIS calibration.  N.M. is supported by JSPS
Research Fellowship for Young Scientists.  This work was supported by
the Grant-in-Aid for the Global COE Program "The Next Generation of
Physics, Spun from Universality and Emergence" from the Ministry of
Education, Culture, Sports, Science and Technology (MEXT) of Japan.


\begin{thebibliography}{}

\bibitem[Balucinska-Church \& McCammon(1992)]{1992ApJ...400..699B} 
Balucinska-Church, M., \& McCammon, D.\ 1992, \apj, 400, 699 

\bibitem[Bamba et al.(2004)]{2004ApJ...602..257B} Bamba, A., Ueno, M., 
Nakajima, H., \& Koyama, K.\ 2004, \apj, 602, 257 

\bibitem[Bautz et al.(2004)]{2004SPIE.5501..111B} Bautz, M.~W., Kissel, 
S.~E., Prigozhin, G.~Y., LaMarr, B., Burke, B.~E., 
\& Gregory, J.~A.\ 2004, \procspie, 5501, 111 

\bibitem[Cash et al.(1980)]{1980ApJ...238L..71C} Cash, W., Charles, P., 
Bowyer, S., Walter, F., Garmire, G., \& Riegler, G.\ 1980, \apjl, 238, L71 

\bibitem[Cooper et al.(2004)]{2004ApJ...605..751C} Cooper, R.~L., Guerrero, 
M.~A., Chu, Y.-H., Chen, C.-H.~R., \& Dunne, B.~C.\ 2004, \apj, 605, 751 

\bibitem[Dennerl et al.(2001)]{2001A&A...365L.202D} Dennerl, K., et al.\ 2001, \aap, 365, L202

\bibitem[Dunne et al.(2001)]{2001ApJS..136..119D} Dunne, B.~C., Points, 
S.~D., \& Chu, Y.-H.\ 2001, \apjs, 136, 119 

\bibitem[Egger 
\& Aschenbach(1995)]{1995A&A...294L..25E} Egger, R.~J., \& Aschenbach, B.\ 1995, \aap, 294, L25 

\bibitem[Guo et al.(1995)]{1995ApJ...453..256G} Guo, Z., Burrows, D.~N., 
Sanders, W.~T., Snowden, S.~L., \& Penprase, B.~E.\ 1995, \apj, 453, 256 

\bibitem[Ishisaki et al.(2007)]{2007PASJ...59S.113I} Ishisaki, Y., et al.\ 
2007, \pasj, 59, 113 

\bibitem[Koyama et al.(1989)]{1989Natur.339..603K} Koyama, K., Awaki, H., 
Kunieda, H., Takano, S., \& Tawara, Y.\ 1989, \nat, 339, 603 

\bibitem[Koyama et al.(2007a)]{2007PASJ...59S..23K} Koyama, K., et al.\ 
2007a, \pasj, 59, 23 

\bibitem[Koyama et al.(2007b)]{2007PASJ...59S.237K} Koyama, K., Uchiyama, 
H., Hyodo, Y., Matsumoto, H., Tsuru, T.~G., Ozaki, M., Maeda, Y., 
\& Murakami, H.\ 2007, \pasj, 59, 237 

\bibitem[Koyama et al.(2007c)]{2007PASJ...59S.245K} Koyama, K., et al.\ 
2007b, \pasj, 59, 245 

\bibitem[Maeda et al.(2002)]{2002ApJ...570..671M} Maeda, Y., et al.\ 2002, 
\apj, 570, 671 

\bibitem[Miller et al.(2008)]{2008PASJ...60S..95D} Miller, E.~D., et al.\ 
2008, \pasj, 60, 95 

\bibitem[Mitsuda et al.(2007)]{2007PASJ...59S...1M} Mitsuda, K., et al.\ 
2007, \pasj, 59, 1

\bibitem[Mori et al.(2008)]{2008PASJ...60S.183M} Mori, H., Tsuru, T.~G., 
Hyodo, Y., Koyama, K., \& Senda, A.\ 2008, \pasj, 60, 183 

\bibitem[Muno et al.(2003)]{2003ApJ...589..225M} Muno, M.~P., et al.\ 2003, 
\apj, 589, 225 

\bibitem[Muno et al.(2006)]{2006ApJS..165..173M} Muno, M.~P., Bauer, F.~E., 
Bandyopadhyay, R.~M., \& Wang, Q.~D.\ 2006, \apjs, 165, 173 

\bibitem[Nakajima et al.(2008)]{2008PASJ...60S...1N} Nakajima, H., et al.\ 
2008, \pasj, 60, 1 

\bibitem[Naz{\'e} et al.(2004)]{2004A&A...418..841N} Naz{\'e}, Y.,
		Antokhin, I.~I., Rauw, G., Chu, Y.-H., Gosset, E., \&
		Vreux, J.-M.\ 2004, \aap, 418, 841 

\bibitem[Plucinsky et al.(1996)]{1996ApJ...463..224P} Plucinsky, P.~P., 
Snowden, S.~L., Aschenbach, B., Egger, R., Edgar, R.~J., 
\& McCammon, D.\ 1996, \apj, 463, 224 

\bibitem[Senda et al.(2003)]{2003ANS...324..151S} Senda, A., Murakami, H., 
\& Koyama, K.\ 2003, Astronomische Nachrichten Supplement, 324, 151 

\bibitem[Serlemitsos et al.(2007)]{2007PASJ...59S...9S} Serlemitsos, P.~J., 
et al.\ 2007, \pasj, 59, 9 

\bibitem[Tawa et al.(2008)]{2008PASJ...60S..11T} Tawa, N., et al.\ 2008, 
\pasj, 60, 11 

\bibitem[Townsley et al.(2006)]{2006AJ....131.2140T} Townsley, L.~K., 
Broos, P.~S., Feigelson, E.~D., Brandl, B.~R., Chu, Y.-H., Garmire, G.~P., 
\& Pavlov, G.~G.\ 2006, \aj, 131, 2140 

\bibitem[Uchiyama et al.(2007)]{2007SPIE.6686E..22U} Uchiyama, H., et al.\ 
2007, \procspie, 6686,  

\bibitem[Uyan{\i}ker et 
al.(2001)]{2001A&A...371..675U} Uyan{\i}ker, B., F{\"u}rst, E., Reich, W., Aschenbach, B., \& Wielebinski, R.\ 2001, \aap, 371, 675 

\bibitem[Yamaguchi et al.(2009)]{2009PASJ...61S.175Y} Yamaguchi, H., Bamba, 
A., \& Koyama, K.\ 2009, \pasj, 61, 175 

\bibitem[Willingale et al.(2003)]{2003MNRAS.343..995W} Willingale, R.,
Hands, A.~D.~P., Warwick, R.~S., Snowden, S.~L., \& Burrows, D.~N.\
2003, \mnras, 343, 995

\end{thebibliography}
\end{document}